\documentclass
[preprint,a4paper,10pt,showpacs,tightenlines,onecolumn]{revtex4-1}%
\usepackage{amsfonts}
\usepackage{amsmath}
\usepackage{amssymb}
\usepackage{graphicx}%
\providecommand{\U}[1]{\protect\rule{.1in}{.1in}}

\begin{document}

\title{Using quantum state protection via dissipation in a quantum-dot molecule to
solve the Deutsch problem }
\author{M. M. Santos, F. O. Prado, H. S. Borges, A. M. Alcalde, J. M. Villas-B\^{o}as,
and E. I. Duzzioni\\Universidade Federal de Uberl\^{a}ndia, Caixa Postal 593, 38400-902,
Uberl\^{a}ndia, MG, Brazil}

\begin{abstract}
The wide set of control parameters and reduced size scale make semiconductor
quantum dots attractive candidates to implement solid-state quantum
computation. Considering an asymmetric double quantum dot coupled by
tunneling, we combine the action of a laser field and the spontaneous emission
of the excitonic state to protect an arbitrary superposition state of the
indirect exciton and ground state. As a by-product we show how to use the
protected state to solve the Deutsch problem.

\end{abstract}
\maketitle

\section{Introduction}

One important feature of Quantum Mechanics is the superposition principle.
This fundamental characteristic for pure states can be summarized by the
quantum entanglement phenomenon \cite{Schrodinger,Horodecki}. It is believed
that such a property can give to the quantum world the advantage of processing
information in a more efficient way than its classical counterpart
\cite{Nielsen 2000}. However, in a realistic scenario, decoherence destroys
the superposition of states leading the system state to a statistical mixture
\cite{Zurek 1981}. A number of methods has been proposed to circumvent this
difficulty; among them we highlight the quantum error-correcting codes
\cite{QECC}, decoherence-free subspaces in collective systems \cite{DFS}, and
dynamical decoupling methods \cite{DD}. Differently from the methods cited
above, the engineering reservoir technique \cite{Poyatos} makes use of
incoherent control of Markovian reservoirs to drive the system state to a
desired superposition of pure states in the asymptotic limit \cite{Kraus
2008}. Therefore, the effective interaction between the system and reservoir
must be carefully engineered to drive the system to equilibrium with the
reservoir. Recently this theme received a great deal of attention in order to
obtain pure entangled protected states \cite{Kraus 2008}. Besides the recent
generalization of the engineering reservoir technique to protect nonstationary
superposition states \cite{Prado 2009, Prado 2011, Xue 2011}, two important
applications of this technique have emerged: (i) the construction of robust
quantum memories \cite{Pastawski 2011} and (ii) the implementation of quantum
computation via dissipation (QCD) \cite{Verstraete 2009}. In the former case
the quantum information can be stored for long times at the nodes of a quantum
network, while in the later case the quantum processor is insensitive to
external perturbations or decoherence. Particularly, in Ref. \cite{Verstraete
2009} it was shown that QCD is universal and can efficiently simulate a given
quantum circuit.

One of the clearest demonstration of the power of quantum processing is the
quantum solution for the Deutsch problem \cite{Deutsch 1985}. Later improved
by Deutsch and Jozsa \cite{DJ} and by Collins, Kim, and Holton \cite{Collins
1998}, the optimized version of the Deutsch algorithm is able to decide wheter
a binary function is constant or balanced with a single measurement, while in
the classical case two measurements are required. The Deutsch problem was
proved using different experimental setups, such as Nuclear Magnetic Resonance
\cite{Jones 1998,Chuang 1998}, optical systems \cite{Mohseni 2003}, circuit
quantum electrodynamics architectures \cite{DiCarlo 2009}, trapped ions
\cite{Gulde 2003}, the nitrogen-vacancy defect center \cite{Shi 2010}, and
quantum dots \cite{Bianucci 2004}.

In this work we investigate the protection of superposition states in a
physical system composed by two asymmetric quantum dots coupled by tunneling
under the application of laser fields, where the interplay of the optical
field and the tunneling creates a structure of two excitonic levels: direct
and indirect, as described in Ref. \cite{Villas-Boas 2004}. The results
obtained here generalize the result obtained in Ref. \cite{Borges 2010}, where
a robust state of the indirect exciton was found. Through the protection of an
arbitrary coherent superposition of the ground and excited (indirect exciton)
states, we propose the implementation of the optimized version of the Deutsch
algorithm \cite{Collins 1998} using QCD. In general, the proposals for
implementing quantum computing protocols make use of coherent control of
quantum systems. In contrast, our proposal to solve the Deutsch problem has
the advantage of being controlled incoherently by the reservoir, becoming
immune to decoherence processes.

The manuscript is divided as follows. In Sec. II we obtain analytically the
protected asymptotic state of the system which is parametrized on the Bloch
sphere. Such a state is used to solve the Deutsch problem in Sec. III.
Conclusions and perspectives for this work are presented in Sec. IV. In the
Appendix we deduce the effective master equation leading to the protected state.

\section{Protected pure state}

We consider two asymmetric quantum dots coupled by tunneling, which under the
application of a static electric field along the growth direction can be
simplified to a three-level system \cite{Villas-Boas 2004}: the ground state
$\left\vert 0\right\rangle $ (no excitation on the system), the exciton state
$\left\vert 1\right\rangle $ (one electron-hole pair in the same dot), and the
indirect exciton state $\left\vert 2\right\rangle $ (the electron in one dot
and the hole in another). The exciton state can be created from the ground
state by the application of a laser field with right the frequency and the
indirect exciton from the exciton state by tunneling of one electron, as
experimentally demonstrated in Ref. \cite{Krenner 2005}. In this way, the
Hamiltonian in the rotating wave approximation for this system is ($\hbar=1$)%
\begin{equation}
H\left(  t\right)  =%
{\displaystyle\sum\limits_{j=0}^{2}}
\omega_{j}\left\vert j\right\rangle \left\langle j\right\vert +T_{e}\left(
\left\vert 1\right\rangle \left\langle 2\right\vert +\left\vert 2\right\rangle
\left\langle 1\right\vert \right)  +\Omega\left[  e^{i\left(  \varphi
+\omega_{L}t\right)  }\left\vert 0\right\rangle \left\langle 1\right\vert
+e^{-i\left(  \varphi+\omega_{L}t\right)  }\left\vert 1\right\rangle
\left\langle 0\right\vert \right]  , \label{hamiltonian}%
\end{equation}
where $\omega_{j}$ is the energy of the $j$th level and $T_{e}$ is the
tunneling coupling between the levels $\left\vert 1\right\rangle $ and
$\left\vert 2\right\rangle $. We tune the gate voltage to have $\omega
_{1}=\omega_{2}=\omega_{L}$ and $\omega_{0}=0$. $\omega_{L}$ and $\varphi$ are
the frequency and phase of the laser field, respectively. $\Omega=\left\langle
0\right\vert \overrightarrow{\mu}\cdot\overrightarrow{E}\left\vert
1\right\rangle $ is the dipole coupling between the excitonic transition, with
$\overrightarrow{\mu}$ being the electric dipole moment and $\overrightarrow
{E}$ the incident electric field of the laser \cite{Note}. According to Ref.
\cite{Borges 2010}, the dominant decoherence processes can effectively be
described by spontaneous emission from states $\left\vert 1\right\rangle $ to
$\left\vert 0\right\rangle $ and $\left\vert 2\right\rangle $ to $\left\vert
0\right\rangle $ with decay rates $\Gamma_{1}$ and $\Gamma_{2}$, respectively.
This approximation remains valid if we consider that (i) the system is at very
low temperatures so that the pure dephasing induced by acoustic phonons does
not modify significantly the system dynamics \cite{Ramsay 2010}; (ii) the
creation of the lower energy exciton state is made by the application of a
resonant laser field, which inhibits transitions mediated by phonons
\cite{Borri 2003, Karwat 2011}; (iii) the optical phonon effects are
negligible because they have quite different frequencies ($\sim30$ meV)
compared with that of our system (see the main text) \cite{Ramsay
2010}.\textit{ }Then, the dynamics of the system can be described by the
master equation%

\begin{equation}
\frac{\partial\rho\left(  t\right)  }{\partial t}=-i\left[  H\left(  t\right)
,\rho\right]  +\mathcal{L}\left(  \rho\right)  , \label{master-eq}%
\end{equation}
with $H\left(  t\right)  $ being the Hamiltonian defined in Eq.
(\ref{hamiltonian}) and $\mathcal{L}\left(  \rho\right)  $ the dissipative
Liouvillian, given by%
\begin{equation}
\mathcal{L}\left(  \rho\right)  =\mathcal{L}_{1}\left(  \rho\right)
+\mathcal{L}_{2}\left(  \rho\right)  =\sum_{i=1}^{2}\frac{\Gamma_{i}}%
{2}\left(  2\left\vert 0\right\rangle \left\langle i\right\vert \rho\left\vert
i\right\rangle \left\langle 0\right\vert -\rho\left\vert i\right\rangle
\left\langle i\right\vert -\left\vert i\right\rangle \left\langle i\right\vert
\rho\right)  . \label{DissLiov}%
\end{equation}

The steady states of the system can be found through the condition%
\begin{equation}
\lim_{t\rightarrow\infty}\frac{\partial\rho\left(  t\right)  }{\partial t}=0,
\label{limite}%
\end{equation}
where $t\rightarrow\infty$ means $t\gg1/\min\Gamma_{i}$. The right-hand side
of Eq. (\ref{master-eq}) fulfills the condition (\ref{limite}) for a pure
state $\left\vert \Psi\right\rangle $ (also called dark state) if
$\mathcal{L}\left(  \left\vert \Psi\right\rangle \left\langle \Psi\right\vert
\right)  =0$ and $H\left\vert \Psi\right\rangle =E\left\vert \Psi\right\rangle
$ \cite{Kraus 2008}. For a master equation in the Lindblad form \cite{Lindblad
1976},%
\[
\mathcal{L}\left(  \rho\right)  =\frac{\gamma}{2}\left(  2O\rho O^{\dag
}-O^{\dag}O\rho-\rho O^{\dag}O\right)  ,
\]
where $\gamma$ is the decay rate and $O$ the jump operator, the state
$\left\vert \Psi\right\rangle $ is the only protected state if $O\left\vert
\Psi\right\rangle =0$ ($\mathcal{L}\left(  \left\vert \Psi\right\rangle
\left\langle \Psi\right\vert \right)  =0$) and there is no further eigenstate
$\left\vert \phi\right\rangle $ of $O$ such that $\left[  O,O^{\dag}\right]
\left\vert \phi\right\rangle =0$ \cite{Carvalho 2001}. The method used here to
obtain the dark state of the system is not unique. For instance, in Refs.
\cite{Cardimona 1982, Hegerfeldt 1992} the dark state conditions are met when
the absorption spectrum is null. Already in Ref. \cite{Swain 2000} a signature
of the dark state is found in the second-order correlation function. In order
to remove the time dependence of the right-hand side of Eq. (\ref{master-eq}),
we move to a rotating frame defined by the unitary transformation
\begin{equation}
U\left(  t\right)  =\exp\left[  \frac{i\omega_{L}t}{2}\left(  \left\vert
1\right\rangle \left\langle 1\right\vert +\left\vert 2\right\rangle
\left\langle 2\right\vert -\left\vert 0\right\rangle \left\langle 0\right\vert
\right)  \right]  . \label{UT}%
\end{equation}
In this frame the Hamiltonian becomes%
\begin{equation}
H_{int}=\Omega\left(  e^{i\varphi}\left\vert 0\right\rangle \left\langle
1\right\vert +e^{-i\varphi}\left\vert 1\right\rangle \left\langle 0\right\vert
\right)  +T_{e}\left(  \left\vert 1\right\rangle \left\langle 2\right\vert
+\left\vert 2\right\rangle \left\langle 1\right\vert \right)  , \label{Hint}%
\end{equation}
the form of $\mathcal{L}\left(  \rho\right)  $ remains unchanged, and $\rho$
is replaced by $\rho_{int}=U^{\dag}\rho U$. Considering the requirements to
obtain a protected pure state, we initially find the eigenvectors of $H_{int}$%
\begin{subequations}
\begin{align}
\left\vert E_{+}\left(  \theta,\varphi\right)  \right\rangle  &  =\frac
{\sin\left(  \theta/2\right)  \left\vert 0\right\rangle +e^{-i\varphi
}\left\vert 1\right\rangle +e^{-i\varphi}\cos\left(  \theta/2\right)
\left\vert 2\right\rangle }{\sqrt{2}},\label{ds1}\\
\left\vert E_{0}\left(  \theta,\varphi\right)  \right\rangle  &  =\cos\left(
\theta/2\right)  \left\vert 0\right\rangle -e^{-i\varphi}\sin\left(
\theta/2\right)  \left\vert 2\right\rangle ,\label{ds2}\\
\left\vert E_{-}\left(  \theta,\varphi\right)  \right\rangle  &  =\frac
{\sin\left(  \theta/2\right)  \left\vert 0\right\rangle -e^{-i\varphi
}\left\vert 1\right\rangle +e^{-i\varphi}\cos\left(  \theta/2\right)
\left\vert 2\right\rangle }{\sqrt{2}}, \label{ds3}%
\end{align}
with eigenvalues $E_{\pm}=\pm\sqrt{\Omega^{2}+T_{e}^{2}}$ and $E_{0}=0$.
$\varphi$ is the laser phase defined above and $\cos\left(  \theta/2\right)
=T_{e}/\sqrt{\Omega^{2}+T_{e}^{2}}$. Through the condition $\mathcal{L}\left(
\left\vert \Psi\right\rangle \left\langle \Psi\right\vert \right)  =0$ we
observe that the dissipative Liouvillians $\mathcal{L}_{1}$ and $\mathcal{L}%
_{2}$ have $\left\{  \left\vert 0\right\rangle ,\left\vert 2\right\rangle
\right\}  $ and $\left\{  \left\vert 0\right\rangle ,\left\vert 1\right\rangle
\right\}  $ as their dark states, respectively. As the relation between the
decay rates is about $\Gamma_{2}=10^{-4}\Gamma_{1}$ with $\Gamma_{1}$ of the
order of $0.33-6.6\mu$eV \cite{Negoita 1999, Chen 2001, Takagahara 2002}, we
conclude that the dissipative dynamics is basically governed by $\mathcal{L}%
_{1}\left(  \rho\right)  $. Note that the eigenvector $\left\vert E_{0}\left(
\theta,\varphi\right)  \right\rangle $ is composed only by $\left\{
\left\vert 0\right\rangle ,\left\vert 2\right\rangle \right\}  $ states.
Therefore, $\left\vert E_{0}\left(  \theta,\varphi\right)  \right\rangle $ is
a dark state of the system, which is in good agreement with our numerical
calculations performed in the regime of parameters defined above. In the
Appendix we analytically show how to obtain the effective master equation
whose protected state is $\left\vert E_{0}\left(  \theta,\varphi\right)
\right\rangle $. Naturally the protected state is not pure; the deviation from
$\left\vert E_{0}\left(  \theta,\varphi\right)  \right\rangle $ introduced by
the indirect exciton decay channel ($\Gamma_{2}$) can be obtained through the
fidelity $\mathcal{F}\left(  \infty\right)  $ in the stationary regime,
defined by%
\end{subequations}
\begin{equation}
\mathcal{F}\left(  \infty\right)  \equiv\lim_{t\rightarrow\infty}\left[
\left\langle E_{0}\left(  \theta,\varphi\right)  \right\vert \rho\left(
t\right)  \left\vert E_{0}\left(  \theta,\varphi\right)  \right\rangle
\right]  =\frac{1+\frac{\Gamma_{2}}{\Gamma_{1}}\frac{T_{e}^{2}}{T_{e}%
^{2}+\Omega^{2}}}{1+\frac{\Gamma_{2}}{\Gamma_{1}}\frac{T_{e}^{4}+2\Omega^{4}%
}{T_{e}^{2}\left(  T_{e}^{2}+\Omega^{2}\right)  }}. \label{fidelity}%
\end{equation}
A simple analysis of the particular case $T_{e}\sim\Omega$ shows that the
fidelity $\mathcal{F}\left(  \infty\right)  \simeq1-\Gamma_{2}/\Gamma_{1}$
attains values next to $1$, even for $\Gamma_{2}$ one order of magnitude lower
than $\Gamma_{1}$.

The state $\left\vert E_{0}\left(  \theta,\varphi\right)  \right\rangle $ can
be represented on the Bloch sphere, where $\theta$ and $\varphi$ are polar and
azimuthal angles. The experimentally accessible values of $\Omega
\simeq0.05-1.0$ meV \cite{Chen 2001, Calarco 2003} and $T_{e}\simeq0.01-10$
meV \cite{Tackeuchi 2000, Emary 2007} enable $\theta$ to vary approximately
from $0.5^{\circ}$ to $179^{\circ}$, while the laser phase $\varphi$ is easily
controlled in the range $[0,2\pi)$. The dependence of $\left\vert E_{0}\left(
\theta,\varphi\right)  \right\rangle $ with respect to $\Omega$ and $T_{e}$ is
analyzed considering three particular cases:

(i) $\Omega\gg T_{e}$: In this case the protected state becomes $\left\vert
E_{0}\left(  \pi,\varphi\right)  \right\rangle =\left\vert 2\right\rangle $
provided that $\theta\rightarrow\pi$. This is achieved increasing the laser
amplitude, since for $\Gamma_{1}\gg\Gamma_{2}$ the lifetime of the indirect
exciton state goes to infinity. This result is in accordance with Ref.
\cite{Borges 2010}.

(ii) $\Omega\ll T_{e}$: In the opposite scenario where $\theta\rightarrow0$
the protected asymptotic state is the ground state of the system $\left\vert
E_{0}\left(  0,\varphi\right)  \right\rangle =\left\vert 0\right\rangle $.
Since the laser amplitude is weak, the direct and indirect exciton states are
not populated.

(iii) $\Omega=T_{e}$: For $\theta=\pi/2$ the protected state $\left\vert
E_{0}\left(  \pi/2,\varphi\right)  \right\rangle =\left(  \left\vert
0\right\rangle -e^{-i\varphi}\left\vert 2\right\rangle \right)  /\sqrt{2}$ is
a coherent superposition of states $\left\vert 0\right\rangle $ and
$\left\vert 2\right\rangle $. In this last case, the control of the laser
relative phase $\varphi$ will enable us to implement the Deustch algorithm, as
shown below.

\section{Deutsch algorithm via dissipative quantum computation}

The Deutsch algorithm was one of the first quantum algorithms to make explicit
use of the quantum parallelism \cite{Deutsch 1985, Collins 1998}. Such an
algorithm was built to decide wheter a given binary function $f:\left\{
0,1\right\}  \rightarrow\left\{  0,1\right\}  $ is constant ($f\left(
0\right)  =f\left(  1\right)  $) or balanced ($f\left(  0\right)  \neq
f\left(  1\right)  $). Differently from the original solution to the Deutsch
problem \cite{Deutsch 1985}, which is probabilistic, we present here a
deterministic algorithm due to Collins, Kim, and Holton \cite{Collins 1998}.
Besides being deterministic, the approach to the Deutsch problem used in Ref.
\cite{Collins 1998} is interesting because only one query to the oracle is
made and auxiliary qubits are unnecessary.

To implement the Deutsch algorithm we use the protected state $\left\vert
E_{0}\left(  \pi/2,\varphi\right)  \right\rangle $ with the laser phase
$\varphi$ being $0$ or $\pi$. In order to clarify the execution of the
algorithm, we make the correspondence between the function domain $\left\{
0,1\right\}  $ and the states of the system $\left\{  \left\vert
0\right\rangle ,\left\vert 2\right\rangle \right\}  $ so that $0\rightarrow
\left\vert 0\right\rangle $ and $1\rightarrow\left\vert 2\right\rangle $. We
define next the parameter $\varepsilon\equiv f\left(  \left\vert
2\right\rangle \right)  -f\left(  \left\vert 0\right\rangle \right)  $, which
can take the values $\left\{  -1,0,1\right\}  $. Therefore, the state
$\left\vert E_{0}\left(  \pi/2,\varphi\right)  \right\rangle $ can be
rewritten as
\begin{equation}
\left\vert E_{0}\left(  \pi/2,\varphi\right)  \right\rangle =\frac{\left\vert
0\right\rangle -e^{-i\pi\varepsilon}\left\vert 2\right\rangle }{\sqrt{2}%
}=\frac{\left\vert 0\right\rangle -\left(  -1\right)  ^{\varepsilon}\left\vert
2\right\rangle }{\sqrt{2}}=\frac{\left\vert 0\right\rangle -\left(  -1\right)
^{f\left(  \left\vert 2\right\rangle \right)  -f\left(  \left\vert
0\right\rangle \right)  }\left\vert 2\right\rangle }{\sqrt{2}}.
\end{equation}
Therefore, if the function is constant ($f\left(  \left\vert 2\right\rangle
\right)  =f\left(  \left\vert 0\right\rangle \right)  $) then $\varphi=0$.
Otherwise, if the function is balanced ($f\left(  \left\vert 2\right\rangle
\right)  \neq f\left(  \left\vert 0\right\rangle \right)  $) then $\varphi
=\pi$. It is assumed that only the oracle has information about the function
$f\left(  i\right)  $. In this case the oracle is the laser phase $\varphi$
programmer. The last part of an algorithm is the readout of the solution. This
is made here, first, by replacing the current laser field by another one with
amplitude $\overline{\Omega}=\Omega\left(  \sqrt{2}+1\right)  $, relative
phase $\overline{\varphi}=0$, and frequency $\overline{\omega}=\omega_{L}$
resonant to the transition $\left\vert 0\right\rangle \rightarrow\left\vert
1\right\rangle $. In this new configuration, $\left\vert E_{0}\left(
\pi/2,\varphi\right)  \right\rangle $ becomes the initial state of system and
the evolved state now is $\overline{\rho}\left(  t\right)  $. In Fig. 1 we
observe the time evolution of the populations $P_{ii}\left(  t\right)
=\left\langle i\right\vert \overline{\rho}\left(  t\right)  \left\vert
i\right\rangle $ of the states $\left\vert 0\right\rangle $ (black solid
line), $\left\vert 1\right\rangle $ (red dashed line), and $\left\vert
2\right\rangle $ (blue dotted line) considering the phases $\varphi=0$ and
$\varphi=\pi$ of the first laser. At specific times $t_{n}=n\pi/\sqrt
{\overline{\Omega}^{2}+T_{e}^{2}}$ with $n=1,3,5,\ldots$, the state of the
system will be $\overline{\rho}\left(  t\right)  \simeq\left\vert
0\right\rangle \left\langle 0\right\vert $ if the phase of the state
$\left\vert E_{0}\left(  \pi/2,\varphi\right)  \right\rangle $ is $\varphi
=\pi$ and $\overline{\rho}\left(  t\right)  \simeq\left\vert 2\right\rangle
\left\langle 2\right\vert $ if the phase is $\varphi=0$. Therefore, applying
another laser pulse resonant with the transition $\left\vert 0\right\rangle
\longleftrightarrow\left\vert 1\right\rangle $ and observing the time-resolved
absorption spectrum \cite{Bianucci 2004} it is possible to distinguish between
the states $\left\vert 0\right\rangle $ and $\left\vert 2\right\rangle $,
provided that if the electron is in state $\left\vert 2\right\rangle $ there
will be no absorption, while if the electron is in state $\left\vert
0\right\rangle $, the light will be absorbed. In summary, discovering wheter
the phase $\varphi$ is $0$ or $\pi$ is equivalent to solving the Deutsch problem.%

\begin{figure}
[h]
\begin{center}
\includegraphics[
height=3.3382in,
width=4.7694in
]%
{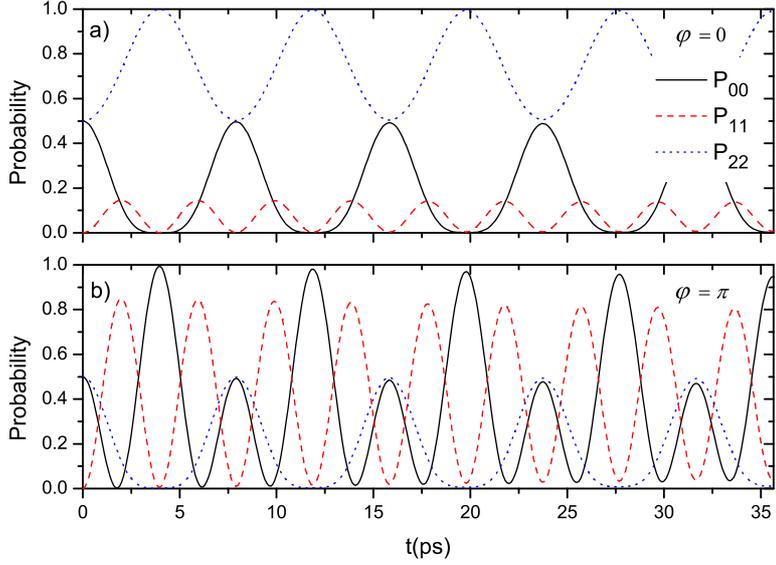}%
\caption{(Color online) Time evolution of the population of the states
$\left\vert 0\right\rangle $ (black solid line), $\left\vert 1\right\rangle $
(red dashed line), and $\left\vert 2\right\rangle $ (blue dotted line)
considering the application of a laser field with amplitude $\overline{\Omega
}=\Omega\left(  \sqrt{2}+1\right)  $, relative phase $\overline{\varphi}=0$,
and frequency $\overline{\omega}=\omega_{L}$. The initial state of the system
is $\left\vert E_{0}\left(  \pi/2,\varphi\right)  \right\rangle $ for (a)
$\varphi=0$ and (b) $\pi$. The physical parameters are $\Omega=T_{e}=200$
$\mu$eV, $\Gamma_{1}=3$ $\mu$eV, and $\Gamma_{2}=10^{-4}\Gamma_{1}$.}%
\end{center}
\end{figure}

To finish our analysis of the Deutsch algorithm, we will show that the error
introduced by the fact that $\left\vert E_{0}\left(  \pi/2,\varphi\right)
\right\rangle $ is not the perfect steady state of the system is negligible.
We define the quantity $\Delta P_{00}\left(  t_{n}\right)  \equiv\left\vert
\left\langle 0\right\vert \overline{\rho}(t_{n},\rho\left(  t\rightarrow
\infty\right)  )-\overline{\rho}\left(  t_{n},\left\vert E_{0}\left(
\pi/2,\varphi\right)  \right\rangle \left\langle E_{0}\left(  \pi
/2,\varphi\right)  \right\vert \right)  \left\vert 0\right\rangle \right\vert
$ as the difference of population in the state $\left\vert 0\right\rangle $
for the density operator $\overline{\rho}\left(  t_{n}\right)  $ considering
two different initial conditions, the approximate state $\left\vert
E_{0}\left(  \pi/2,\varphi\right)  \right\rangle $ and the exact state
$\rho\left(  t\rightarrow\infty\right)  $, where the later is obtained by
numerical calculation. Tables I and II show the dependence of $\Delta
P_{00}\left(  t_{n}\right)  $ on the ratio $\epsilon\equiv\Gamma_{2}%
/\Gamma_{1}$ for $\varphi=0$ and $\pi$, respectively.\begin{table}[h]
\caption{The dependence of $\Delta P_{00}(t_{n})$ on the ratio $\epsilon
=\Gamma_{2}/\Gamma_{1}$ for $\varphi=0$. }%
\begin{ruledtabular}
\begin{tabular}{llll}
& $\epsilon = 0.0001$ & $\epsilon = 0.001$ & $\epsilon = 0.01$\\
\cline{2-4}
$\Delta P_{00}(t_{1})$ & $-4.9913\times10^{-5}$ & $-4.9847\times10^{-4}$ & $-4.9178\times10^{-3}$\\
$\Delta P_{00}(t_{2})$ & $-4.9757\times10^{-5}$ & $-4.9689\times10^{-4}$ & $-4.9010\times10^{-3}$\\
$\Delta P_{00}(t_{3})$ & $-4.9597\times10^{-5}$ & $-4.9527\times10^{-4}$ & $-4.8838\times10^{-3}$\\
$\Delta P_{00}(t_{4})$ & $-4.9434\times10^{-5}$ & $-4.9363\times10^{-4}$ & $-4.8665\times10^{-3}$\\
\end{tabular}
\end{ruledtabular}\end{table}\pagebreak

\begin{table}[h]
\caption{The dependence of $\Delta P_{00}(t_{n})$ on the ratio $\epsilon
=\Gamma_{2}/\Gamma_{1}$ for $\varphi=\pi$. }%
\begin{ruledtabular}
\begin{tabular}{llll}
& $\epsilon = 0.0001$ & $\epsilon = 0.001$ & $\epsilon = 0.01$\\
\cline{2-4}
$\Delta P_{00}(t_1)$ & $9.9108\times10^{-5}$ & $9.8974\times10^{-4}$ & $9.7653\times10^{-3}$\\
$\Delta P_{00}(t_2)$ & $9.7382\times10^{-5}$ & $9.7249\times10^{-4}$ & $9.5942\times10^{-3}$\\
$\Delta P_{00}(t_3)$ & $9.5694\times10^{-5}$ & $9.5563\times10^{-4}$ & $9.4268\times10^{-3}$\\
$\Delta P_{00}(t_4)$ & $9.4043\times10^{-5}$ & $9.3913\times10^{-4}$ & $9.2361\times10^{-3}$\\
\end{tabular}
\end{ruledtabular}\end{table}

\newpage

\section{Conclusions and perspectives}

In summary, we showed the existence of a pure protected state in a system
composed by two asymmetric quantum dots coupled by tunneling and driven by a
laser field. The direct exciton state is under strong spontaneous decay. The
asymptotic protected state is a superposition of the ground and the indirect
exciton state. By controlling the ratio between the laser amplitude and
tunneling rate it is possible to control the polar angle, while controlling
the laser phase enables the control of the azimuthal angle of the Bloch
sphere. The scheme for state protection remains true at low temperatures
($T\sim0$ K). For high temperatures the phonon contribution becomes important
so that the phonon-induced dephasing will destroy the superposition of the
protected state. As an application of dissipative quantum computation we
proposed the implementation of the Deutsch algorithm in this system, which
basically consists of distinguishing the relative phase $0$ or $\pi$ between
the states $\left\vert 0\right\rangle $ and $\left\vert 2\right\rangle $.
Differently from the usual proposals for implementing quantum algorithms,
which are based on unitary evolutions, here we make use of incoherent
evolution of the Markovian reservoir. This approach is interesting because it
is naturally immune to external perturbations and decoherence processes. A
similar scheme might protect the state of two or more qubits which can be used
to implement more sophisticated algorithms, such as Deutsch-Jozsa and Grover algorithms.

\appendix*

\section{Effective master equation}

For the regime of parameters considered here, the master equation
(\ref{master-eq}) has an analytical solution. However, the method developed in
this appendix can be useful to obtain the effective master equation for a more
elaborate problem. First we rewrite Eq. (\ref{master-eq}) in the interaction
picture according to the unitary transformation (\ref{UT}):%
\begin{equation}
\frac{\partial\rho_{int}\left(  t\right)  }{\partial t}=-i\left[  H_{int}%
,\rho_{int}\right]  +\mathcal{L}_{int}\left(  \rho_{int}\right)  ,
\label{MasterEqA}%
\end{equation}
where $H_{int}$ is given by Eq. (\ref{Hint}) and the dissipative Liouvillian
by Eq. (\ref{DissLiov}) with $\rho$ replaced by $\rho_{int}$. Performing a
change of basis to the Hamiltonian eigenstates, $H_{int}$ in Eq.
(\ref{MasterEqA}) can be expressed as%
\begin{equation}
H_{int}=E_{+}\left(  \left\vert E_{+}\right\rangle \left\langle E_{+}%
\right\vert -\left\vert E_{-}\right\rangle \left\langle E_{-}\right\vert
\right)  ,
\end{equation}
and the dissipative Liouvillians $\mathcal{L}_{int}\left(  \rho_{int}\right)
=\mathcal{L}_{1}\left(  \rho_{int}\right)  +\mathcal{L}_{2}\left(  \rho
_{int}\right)  $ as \begin{widetext}%
\begin{eqnarray}
\mathcal{L}_{1}\left( \rho_{int}\right)  &= &   \nonumber \\
& &\frac{\Gamma_{1}}{4}
\left[ \sin^{2}\left(  \theta/2\right)
\left[
\sqrt
{2}\cot\left(  \theta/2\right)  \left(  \left\vert E_{0}\right\rangle
\left\langle E_{+}\right\vert -\left\vert E_{0}\right\rangle \left\langle
E_{-}\right\vert \right)  +\left\vert E_{+}\right\rangle \left\langle
E_{+}\right\vert -\left\vert E_{-}\right\rangle \left\langle E_{-}\right\vert
+\left\vert E_{-}\right\rangle \left\langle E_{+}\right\vert -\left\vert
E_{+}\right\rangle \left\langle E_{-}\right\vert \right]  \rho_{int} \right. \nonumber
\\
& &     \left. \times\left[  \sqrt{2}\cot\left(  \theta/2\right)  \left(  \left\vert
E_{+}\right\rangle \left\langle E_{0}\right\vert -\left\vert E_{-}%
\right\rangle \left\langle E_{0}\right\vert \right)  +\left\vert
E_{+}\right\rangle \left\langle E_{+}\right\vert -\left\vert E_{-}%
\right\rangle \left\langle E_{-}\right\vert -\left\vert E_{-}\right\rangle
\left\langle E_{+}\right\vert +\left\vert E_{+}\right\rangle \left\langle
E_{-}\right\vert \right]  \right.  \nonumber \\
& & \left.  -\left\{  \rho_{int},\left\vert E_{+}\right\rangle \left\langle
E_{+}\right\vert +\left\vert E_{-}\right\rangle \left\langle E_{-}\right\vert
-\left\vert E_{-}\right\rangle \left\langle E_{+}\right\vert -\left\vert
E_{+}\right\rangle \left\langle E_{-}\right\vert \right\}  \right]
\end{eqnarray}
and%
\begin{eqnarray}
\mathcal{L}_{2}\left(  \rho_{int}\right) & = & \frac{\Gamma_{2}}{4}\left[
\sin^{2}\left(  \theta/2\right)  \cos^{2}\left(  \theta/2\right)  \left[
\sqrt{2}\cot\left(  \theta/2\right)  \left(  \left\vert E_{0}\right\rangle
\left\langle E_{-}\right\vert -\left\vert E_{-}\right\rangle \left\langle
E_{0}\right\vert -\left\vert E_{+}\right\rangle \left\langle E_{0}\right\vert
+\left\vert E_{0}\right\rangle \left\langle E_{+}\right\vert \right)  \right.
\right.  \nonumber \\
& & +\left.  \left\vert E_{+}\right\rangle \left\langle E_{+}\right\vert
+\left\vert E_{-}\right\rangle \left\langle E_{-}\right\vert -2\left\vert
E_{0}\right\rangle \left\langle E_{0}\right\vert +\left\vert E_{-}%
\right\rangle \left\langle E_{+}\right\vert +\left\vert E_{+}\right\rangle
\left\langle E_{-}\right\vert \right]  \rho_{int} \nonumber \\
& & \times\left[  -\sqrt{2}\cot\left(  \theta/2\right)  \left(  \left\vert
E_{0}\right\rangle \left\langle E_{-}\right\vert -\left\vert E_{-}%
\right\rangle \left\langle E_{0}\right\vert -\left\vert E_{+}\right\rangle
\left\langle E_{0}\right\vert +\left\vert E_{0}\right\rangle \left\langle
E_{+}\right\vert \right)  \right.  \nonumber \\
& & +\left.  \left\vert E_{+}\right\rangle \left\langle E_{+}\right\vert
+\left\vert E_{-}\right\rangle \left\langle E_{-}\right\vert -2\left\vert
E_{0}\right\rangle \left\langle E_{0}\right\vert +\left\vert E_{-}%
\right\rangle \left\langle E_{+}\right\vert +\left\vert E_{+}\right\rangle
\left\langle E_{-}\right\vert \right] \nonumber \\
& & -\cos^{2}\left(  \theta/2\right)  \left\{  \rho_{int},\left\vert
E_{+}\right\rangle \left\langle E_{+}\right\vert +\left\vert E_{-}%
\right\rangle \left\langle E_{-}\right\vert +\left\vert E_{-}\right\rangle
\left\langle E_{+}\right\vert +\left\vert E_{+}\right\rangle \left\langle
E_{-}\right\vert \right. \nonumber  \\
& & \left.  \left.  -\sqrt{2}\tan\left(  \theta/2\right)  \left(  \left\vert
E_{0}\right\rangle \left\langle E_{-}\right\vert +\left\vert E_{-}%
\right\rangle \left\langle E_{0}\right\vert +\left\vert E_{+}\right\rangle
\left\langle E_{0}\right\vert +\left\vert E_{0}\right\rangle \left\langle
E_{+}\right\vert \right)  +2\tan^{2}\left(  \theta/2\right)  \left\vert
E_{0}\right\rangle \left\langle E_{0}\right\vert \right\}  \right],
\end{eqnarray}
\end{widetext}where $\left\{  a,b\right\}  =ab+ba$ states for the
anticommutator. The action of a unitary transformation $\widetilde{U}\left(
t\right)  =\exp\left(  -iH_{int}t\right)  $ on Eq. (\ref{MasterEqA}) is able
to remove the unitary part of its dynamics. Such procedure is interesting
because the operators of the form $\left\vert E_{i}\right\rangle \left\langle
E_{j}\right\vert $ with $i,j=\{+,-,0\}$ for $i\neq j$ in Eq. (\ref{MasterEqA})
will oscillate quickly as shown here:%
\begin{subequations}
\begin{align}
\widetilde{U}^{\dag}\left(  t\right)  \left\vert E_{+}\right\rangle
\left\langle E_{0}\right\vert \widetilde{U}\left(  t\right)   &  =\left\vert
E_{+}\right\rangle \left\langle E_{0}\right\vert e^{iE_{+}t},\\
\widetilde{U}^{\dag}\left(  t\right)  \left\vert E_{-}\right\rangle
\left\langle E_{0}\right\vert \widetilde{U}\left(  t\right)   &  =\left\vert
E_{-}\right\rangle \left\langle E_{0}\right\vert e^{-iE_{+}t},\\
\widetilde{U}^{\dag}\left(  t\right)  \left\vert E_{+}\right\rangle
\left\langle E_{-}\right\vert \widetilde{U}\left(  t\right)   &  =\left\vert
E_{+}\right\rangle \left\langle E_{-}\right\vert e^{2iE_{+}t}.
\end{align}
Since $\Gamma_{2}\ll\Gamma_{1}\ll\Omega,T_{e}$ it is possible to perform the
rotating wave approximation, leading to an effective master equation%
\end{subequations}
\begin{equation}
\frac{\partial\rho_{eff}\left(  t\right)  }{\partial t}=\mathcal{L}%
_{eff}\left(  \rho_{eff}\right)  .
\end{equation}
The dissipative Liouvillian $\mathcal{L}_{eff}\left(  \rho_{eff}\right)  $ can
be written in the Lindblad form as\begin{widetext}
\begin{equation}
\mathcal{L}_{eff}\left(  \rho_{eff}\right)  =\sum_{\alpha=1}^{2}\sum_{i=1}
^{5}\frac{\Gamma_{\alpha,i}}{2}\left(  2O_{\alpha,i}\rho_{eff}O_{\alpha
,i}^{\dag}-O_{\alpha,i}^{\dag}O_{\alpha,i}\rho_{eff}-\rho_{eff}O_{\alpha
,i}^{\dag}O_{\alpha,i}\right),  \label{meqf}\\
\end{equation}
where
\begin{center}
\begin{tabular}
[c]{ll}
$\Gamma_{1,1}=\frac{\Gamma_{1}}{8}\sin^{2}\left(  \theta/2\right)  $ &
$O_{1,1}=\left\vert E_{+}\right\rangle\left\langle E_{+}\right\vert-\left\vert
E_{-}\right\rangle\left\langle E_{-}\right\vert$\\
$\Gamma_{1,2}=\Gamma_{1,1}$ & $O_{1,2}=\left\vert E_{-}\right\rangle
\left\langle E_{+}\right\vert$\\
$\Gamma_{1,3}=\Gamma_{1,1}$ & $O_{1,3}=\left\vert E_{+}\right\rangle
\left\langle E_{-}\right\vert$\\
$\Gamma_{1,4}=\frac{\Gamma_{1}}{4}\cos^{2}\left(  \theta/2\right)  $ &
$O_{1,4}=\left\vert E_{0}\right\rangle\left\langle E_{-}\right\vert$\\
$\Gamma_{1,5}=\Gamma_{1,4}$ & $O_{1,5}=\left\vert E_{0}\right\rangle
\left\langle E_{+}\right\vert$\\
$\Gamma_{2,1}=\frac{\Gamma_{2}}{8}\sin^{2}\left(  \theta/2\right)  \cos
^{2}\left(  \theta/2\right)  \qquad$ & $O_{2,1}=\left\vert E_{+}\right
\rangle\left\langle E_{+}\right\vert-2\left\vert E_{0}\right\rangle
\left\langle E_{0}\right\vert+\left\vert E_{-}\right\rangle\left\langle
E_{-}\right\vert$\\
$\Gamma_{2,2}=\Gamma_{2,1}$ & $O_{2,2}=\left\vert E_{-}\right\rangle
\left\langle E_{+}\right\vert$\\
$\Gamma_{2,3}=\Gamma_{2,1}$ & $O_{2,3}=\left\vert E_{+}\right\rangle
\left\langle E_{-}\right\vert$\\
$\Gamma_{2,4}=2\Gamma_{2,1}$ & $O_{2,4}=\cot\left(  \theta/2\right)
\left\vert E_{0}\right\rangle\left\langle E_{+}\right\vert-\tan\left(
\theta/2\right)  \left\vert E_{-}\right\rangle\left\langle E_{0}\right\vert$\\
$\Gamma_{2,5}=2\Gamma_{2,1}$ & $O_{2,5}=\cot\left(  \theta/2\right)
\left\vert E_{0}\right\rangle\left\langle E_{-}\right\vert-\tan\left(
\theta/2\right)  \left\vert E_{+}\right\rangle\left\langle E_{0}\right\vert.$
\end{tabular}
\end{center}
\end{widetext}

\bigskip

\noindent In Eq. (\ref{meqf}), the index $\alpha$ refers to the direct
($\Gamma_{1}$) and indirect ($\Gamma_{2}$) exciton decay rates and index $i$
enumerates the operators. From this effective master equation it is easy to
see that $\left\vert E_{0}\right\rangle $ is the only eigenstate with null
eigenvalue of operators $O_{1,i}$. Although the operators $O_{2,i}$ presented
in $\mathcal{L}_{eff}\left(  \rho_{eff}\right)  $ rotate the state $\left\vert
E_{0}\right\rangle $, the decay rates $\Gamma_{2,i}$ are much lower than
$\Gamma_{1,i}$. In summary, the dissipative dynamics is dictated by the direct
exciton decay, ensuring that the only protected pure state is $\left\vert
E_{0}\right\rangle $. To prove the uniqueness of the protected state, we solve
Eq. (\ref{MasterEqA}) using the software Wolfram \textsc{Mathematica} 7 for an
arbitrary initial state with density matrix elements $\rho_{ij}\left(
0\right)  $ with $i,j=0,1,2$ so that $\sum\limits_{i=0}^{2}\rho_{ii}\left(
0\right)  =1$. We get the expression for the density matrix elements in the
asymptotic time ($t\rightarrow\infty$):%
\begin{align*}
\rho_{00}\left(  \infty\right)   &  =\frac{T_{e}^{4}\left(  T_{e}^{2}%
+\Omega^{2}\right)  \Gamma_{1}+\left(  T_{e}^{6}+\Omega^{6}\right)  \Gamma
_{2}}{\left(  T_{e}^{2}+\Omega^{2}\right)  \left[  T_{e}^{2}\left(  T_{e}%
^{2}+\Omega^{2}\right)  \Gamma_{1}+\left(  T_{e}^{4}+2\Omega^{4}\right)
\Gamma_{2}\right]  },\\
\rho_{11}\left(  \infty\right)   &  =\frac{\Omega^{4}\Gamma_{2}}{T_{e}%
^{2}\left(  T_{e}^{2}+\Omega^{2}\right)  \Gamma_{1}+\left(  T_{e}^{4}%
+2\Omega^{4}\right)  \Gamma_{2}},\\
\rho_{22}\left(  \infty\right)   &  =\frac{T_{e}^{2}\Omega^{2}\left(
T_{e}^{2}+\Omega^{2}\right)  \left(  \Gamma_{1}+\Gamma_{2}\right)  }{\left(
T_{e}^{2}+\Omega^{2}\right)  \left[  T_{e}^{2}\left(  T_{e}^{2}+\Omega
^{2}\right)  \Gamma_{1}+\left(  T_{e}^{4}+2\Omega^{4}\right)  \Gamma
_{2}\right]  },\\
\rho_{02}\left(  \infty\right)   &  =-e^{i\varphi}T_{e}\Omega\frac{T_{e}%
^{2}\left(  T_{e}^{2}+\Omega^{2}\right)  \Gamma_{1}+\left(  T_{e}^{4}%
-\Omega^{4}\right)  \Gamma_{2}}{\left(  T_{e}^{2}+\Omega^{2}\right)  \left[
T_{e}^{2}\left(  T_{e}^{2}+\Omega^{2}\right)  \Gamma_{1}+\left(  T_{e}%
^{4}+2\Omega^{4}\right)  \Gamma_{2}\right]  },\\
\rho_{01}\left(  \infty\right)   &  =\rho_{12}\left(  \infty\right)  =0.
\end{align*}
Note that the elements $\rho_{ij}\left(  \infty\right)  $ are independent of
$\rho_{ij}\left(  0\right)  $. Considering the range of values for the
parameters $\Gamma_{1}$, $\Gamma_{2}$, $T_{e}$, and $\Omega$ used in the
paper, $\Gamma_{1}\gg\Gamma_{2}$ and $T_{e}\geq\Omega$, the matrix element
$\rho_{11}\left(  \infty\right)  \ll\rho_{00}\left(  \infty\right)  ,\rho
_{02}\left(  \infty\right)  ,\rho_{22}\left(  \infty\right)  $. Then, the
protected state is approximately $\left\vert E_{0}\left(  \theta
,\varphi\right)  \right\rangle $ independently of the initial state of the
system. Another way to prove the uniqueness of the protected state $\left\vert
E_{0}\left(  \theta,\varphi\right)  \right\rangle $ is using theorem 2 of Ref.
\cite{Kraus 2008}. Defining the jump operator $c\equiv\left\vert
0\right\rangle \left\langle 1\right\vert $ and the "subspace" $S\equiv\left\{
\gamma\left\vert E_{-}\left(  \theta,\varphi\right)  \right\rangle
+\delta\left\vert E_{+}\left(  \theta,\varphi\right)  \right\rangle \text{
with }\gamma,\delta\in%
\mathbb{C}
\text{ }\right.  $ $\left.  \text{and }\left\vert \gamma\right\vert
^{2}+\left\vert \delta\right\vert ^{2}=1\right\}  $, we observe that $S\neq
cS$ for every $\gamma$, $\delta$, $\theta$, and $\varphi$.

The time scale necessary to the system state to be stationary is
$t_{ss}=1/\min\Gamma_{1,i}$ ($i=1,\ldots,5$), which depends on $\Gamma_{1}$,
$T_{e}$, and $\Omega$. Returning to the Schr\"{o}dinger picture, we observe
that the unitary transformation $\widetilde{U}\left(  t\right)  $ does not
affect the protected state, while $U\left(  t\right)  $ introduces a relative
phase between the states $\left\vert 0\right\rangle $ and $\left\vert
2\right\rangle $.

\begin{acknowledgments}
The authors would like to thank CAPES, FAPEMIG, CNPq, and the Brazilian
National Institutes of Science and Technology for Quantum Information
(INCT-IQ) and for Semiconductor Nanodevices (INCT-DISSE) for financial support.
\end{acknowledgments}


\begin{thebibliography}{99}
\bibitem {Note}The dipole coupling between the ground state and the indirect
exciton is very small and we will neglect it here.
\end{thebibliography}
\end{document}